\documentstyle[11pt,aaspp4,flushrt]{article}

\newcommand\gsim{\mathrel{\raise.3ex\hbox{$>$}\mkern-14mu
             \lower0.6ex\hbox{$\sim$}}}
\newcommand\lsim{\mathrel{\raise.3ex\hbox{$<$}\mkern-14mu
             \lower0.6ex\hbox{$\sim$}}}

\input epsf

\def\figurepsxy[#1,#2,#3]#4.{\midinsert\parindent=0pt\eightpoint
    \vbox{\epsfxsize=#2\epsfysize=#3\centerline{\epsfbox{#1}}}
    \def\par{\endgraf\endinsert}{\bf Figure#4.}}
\def\arcs{\ifmmode {^{\scriptscriptstyle\prime\prime}}
          \else $^{\scriptscriptstyle\prime\prime}$\fi}
\def\arcm{\ifmmode {^{\scriptscriptstyle\prime}}
          \else $^{\scriptscriptstyle\prime}$\fi}
\newdimen\sa  \newdimen\sb
\def\parcs{\sa=.07em \sb=.03em
     \ifmmode $\rlap{.}$^{\scriptscriptstyle\prime\kern -\sb\prime}$\kern -\sa$
     \else \rlap{.}$^{\scriptscriptstyle\prime\kern -\sb\prime}$\kern -\sa\fi}
\def\parcm{\sa=.08em \sb=.03em
     \ifmmode $\rlap{.}\kern\sa$^{\scriptscriptstyle\prime}$\kern-\sb$
     \else \rlap{.}\kern\sa$^{\scriptscriptstyle\prime}$\kern-\sb\fi}

\begin{document}

\title{TWISTING OF X-RAY ISOPHOTES IN TRIAXIAL GALAXIES}
\author{Aaron J. Romanowsky \\
  Christopher S. Kochanek}
\affil{Harvard-Smithsonian Center for Astrophysics, MS-10, \\
       60 Garden Street,
       Cambridge MA 02138 \\
       Email: aromanowsky@cfa.harvard.edu}
\authoraddr{MS-10 \\ 
       60 Garden Street \\
       Cambridge MA 02138  \protect \\
       Email: aromanowsky@cfa.harvard.edu}
 
\authoremail{aromanowsky@cfa.harvard.edu}
\authoremail{ckochanek@cfa.harvard.edu}

\begin{abstract}
We investigate X-ray isophote twists created by triaxiality differences between 
the luminous stellar distributions and the dark halos in elliptical galaxies.
For a typically oblate luminous galaxy embedded in a more prolate halo formed by 
dissipationless collapse, the triaxiality difference of $\Delta T \simeq 0.7$ 
leads to typical isophote twists of 
$\langle \Delta \psi_{\rm X} \rangle \simeq 16^\circ \pm 19^\circ$
at 3 stellar effective radii. In a model which includes baryonic dissipation
the effect is smaller, with $\Delta T \simeq 0.3$ and 
$\langle \Delta \psi_{\rm X} \rangle \simeq 5^\circ \pm 8^\circ$.
Thus, accurate measurements of X-ray isophote twists may be able to set 
constraints on the interactions between baryons and dissipationless dark matter 
during galaxy formation. The 30$^\circ$ X-ray isophote twist in the E4 galaxy 
NGC 720 cannot be reproduced by our model, suggesting an intrinsic misalignment 
between the halo and the stars rather than a projection effect.
\end{abstract}

\section{INTRODUCTION}
Determining the intrinsic shapes of the luminosity and mass distributions of
elliptical galaxies is an important unresolved problem. Both observational 
constraints from individual systems and predictions from theory demonstrate 
that the stellar and dark matter components of ellipticals are well described 
as triaxial ellipsoids. Parametric (e.g., Ryden 1992; Lambas, Maddox, \& Loveday 
1992) and non-parametric (Fasano \& Vio 1991; Tremblay \& Merritt 1995; 
Ryden 1996) inversions of the ellipticity distribution of the stellar components
rule out the hypotheses that elliptical galaxies are all oblate or all prolate. 
Because of the information lost in projection, inversions that include 
triaxiality cannot determine a unique shape distribution, but the ellipticals 
seem to fall into two classes: bright ($M_B < -20$),  triaxial, boxy galaxies 
and faint, roughly oblate, disky galaxies (Tremblay \& Merritt 1996). This 
division is consistent with other structural, dynamical, and evolutionary 
evidence for two classes of ellipticals (e.g., Fasano 1991; Fasano \& Vio 1991; 
Busarello, Longo, \& Feoli 1992; Capaccioli, Caon, \& D'Onofrio 1992; Nieto, 
Poulain, \& Davoust 1994; J{\o}rgensen \& Franx 1994; Kormendy \& Bender 1996).

Binney (1978, 1985; see also Contopoulos 1956) first noted that a triaxial 
galaxy with its net angular momentum vector aligned with its short axis would 
show a ``kinematic misalignment'' between the projected axes of the rotation 
and the light, and surveys of elliptical galaxy kinematics have found that such 
kinematic misalignments are common (Davies \& Birkinshaw 1988; Franx, 
Illingworth, \& Heckman 1989; Jedrzejewski \& Schechter 1989). However, a 
kinematic misalignment may also be caused by an {\it intrinsic} misalignment 
between the rotational and the short axes because the angular momentum vector of 
an equilibrium triaxial system can lie anywhere in the plane of the long and 
short axes (Heiligman \& Schwarzschild 1979). Alternatively, the galaxy may not 
be in equilibrium (due to a recent accretion event or to tidal forces from a 
satellite galaxy), or it may be a misidentified S0 with a bar (see Merritt 1997a 
for a review of examples). A statistical analysis of kinematic misalignments in 
bright ellipticals (Franx, Illingworth, \& de Zeeuw 1991) found limits on the 
mean stellar triaxiality and the mean intrinsic misalignment angle of
$\langle T_* \rangle \leq$ 0.7 and 
$\langle \psi_{\rm int} \rangle \leq$ 45$^\circ$, where the upper bounds for 
both parameters occurred only for solutions with two widely-separated peaks.
Although degeneracies prevented strong constraints on $T_*$, there were 
indications for a bimodal population of galaxies, with a large fraction of 
nearly-oblate, short-axis rotators, and a small fraction of nearly-prolate, 
long-axis rotators. Tenjes et al. (1993) and Statler (1994a,b, Statler \& Fry 
1994) have devised methods which invert the projected streamlines of stellar 
orbits to determine the intrinsic shape of a galaxy, and their initial results 
also indicate the existence of a bimodal population (Statler, Dutta, \& Bak 
1997; Bak \& Statler 1997).

Theoretical models of halo formation generally produce flattened, 
prolate-triaxial halos (e.g., Bardeen et al. 1986; Frenk et al. 1988; White \& 
Ostriker 1990; Aguilar \& Merritt 1990; Katz 1991; Cannizzo \& Hollister 1992;
Cole \& Lacey 1996). For example, $N$-body simulations of a small sample of 
halos by Dubinski (1991, 1992) and Dubinski \& Carlberg (1991) found that at 
25 kpc they are very flat (mean short axis ratio of 
$\langle c_{\rm d} \rangle \simeq 0.42 \pm 0.06$) and nearly-prolate
($\langle T_{\rm d} \rangle \simeq 0.8 \pm 0.2$). The final halo shapes and 
angular momenta were very sensitive to the initial cosmological tidal field,
and the angular momentum vectors were well-aligned with the short axes, with 
an average intrinsic misalignment of 
$\langle \psi_{\rm int} \rangle \simeq 26^\circ \pm 22^\circ$. Warren et al. 
(1992) found similar results ($\langle c_{\rm d} \rangle \simeq 0.62 \pm 0.13$
and $\langle T_{\rm d} \rangle \simeq 0.68 \pm 0.24$ at 40 kpc), and found 
that the more massive halos were slightly flatter and more prolate. Again, the 
angular momentum vector was most commonly aligned with the short axis. We
expect dissipative processes to produce baryonic galaxies with shapes that are 
drastically different from those of their parent halos (e.g., 
Katz \& Gunn 1991; Udry 1993; Navarro \& White 1994), but the concomitant 
effects on the shapes of the halos are unclear. Preliminary simulations 
suggest that halos become rounder and more oblate through interactions with 
baryons (Evrard, Summers, \& Davis 1994; Dubinski 1994), probably due to the 
destruction and damping of halo box orbits.  Dubinski (1994) found that the 
flatness and prolateness are reduced relative to a dissipationless model
($c_{\rm d} \sim$ 0.6 versus 0.4, and $T_{\rm d} \sim$ 0.4-0.5 versus 0.8, at 
20 kpc).  Late, major mergers of disk galaxies in compact groups may have
produced some fraction of the present population of ellipticals (e.g., Heyl, 
Hernquist, \& Spergel 1994; Weil \& Hernquist 1996; Barnes \& Hernquist 1996).
Although the simulations have not carefully explored the effects of the 
mergers on the shapes of the dark halos, Weil \& Hernquist (1996) suggest that 
the shapes of remnant halos are rather round and oblate-triaxial, having no 
correlation with the shapes of the remnant stellar distributions. Current 
theoretical and observational results thus suggest that the modestly 
triaxial-oblate luminous parts of elliptical galaxies are embedded in halos 
which are more prolate, so that there should be a strong gradient with radius 
in the triaxiality of the mass distribution.

Direct measurements of the shapes of halos are far more difficult than 
measurements of the shapes of the stellar distributions.  Careful examination 
of the kinematics of polar rings (Arnaboldi et al. 1993), gas disks 
(Lees 1991; Bertola et al. 1991; Franx, van Gorkom, \& de Zeeuw 1994; Plana 
\& Boulesteix 1996; Bureau \& Freeman 1997), and planetary nebulae (Hui et 
al. 1995; Mathieu, Dejonghe, \& Hui 1996) have been used to constrain the 
shapes of both stellar and dark matter potentials.  Unfortunately, such 
tracers are rarely found at large enough galactocentric radii to be helpful 
for probing the dark matter potential, and it is likely that there are 
systematic correlations between the shapes of halos and the presence of rings 
and disks.  Gravitational lenses can also be used to directly measure the 
shapes of mass distributions, and the data appear to require very flat mass 
distributions. However, the number of lenses is small, and the quantitative 
effects of external tidal perturbations are not yet understood (King \& 
Browne 1996; Kochanek 1996; Keeton, Kochanek, \& Seljak 1997; Witt \& Mao 
1997). The most promising candidate for directly measuring the intrinsic 
shapes of nearby halos is high-resolution mapping of the X-ray emission from 
hot gas in the halo potential. For example, Buote \& Canizares (1994, 1996b, 
1997; hereafter BC94, BC96b, BC97) examined the radial profiles, position 
angles, and axis ratios of the X-ray isophotes of the bright, isolated E4 
galaxy NGC 720, and found that the shape and the position angle of the 
potential cannot be produced by the stars. Thus, by using only geometric 
evidence and the assumption that the gas is in quasi-hydrostatic equilibrium, 
they showed that the stars must be embedded in a more massive, {\it flatter}, 
dark matter halo.  For the S0 galaxy NGC 1332, Buote \& Canizares (1996a) 
reached a qualitatively similar conclusion.

X-ray observations can also be used to constrain the triaxialities of halos 
by searching for X-ray isophote twists. The projection of a triaxial light 
distribution whose triaxiality varies with radius produces isophotes whose 
axis orientations vary with radius (see Mihalas \& Binney 1981). Isophote 
twists in the optical surface brightness are common (e.g., Williams \& 
Schwarzschild 1979; Leach 1981), and can be used to constrain the shape of 
an individual galaxy (Fasano 1995), although care must be taken to rule out 
intrinsic axis twist caused by other factors (Fasano \& Bonoli 1989; Nieto 
et al. 1992). X-ray emission from a triaxial system can show an analogous 
isophote twist --- as pointed out by Binney (1978) for the case of a galaxy 
cluster --- allowing the triaxiality of an individual galaxy's potential to 
be probed. In the simplest case, if a galaxy's luminous and dark matter 
distributions are intrinsically aligned and have {\it constant} but 
{\it different} triaxialities, then the projected axes of its X-ray 
isophotes will twist from small radii, where the potential is dominated by 
the stellar core, to large radii, where it is dominated by the dark halo.
Similarly, as discussed by BC96b, the radial triaxiality gradients produced
in halos by dissipative processes can result in X-ray isophote twists.
Here we employ simple models to study the behavior of X-ray twists, and to 
determine how well they constrain the triaxiality of halos. In \S 2.1, we 
use an analytic approximation to find the amplitudes of twists expected for 
different assumptions about the shapes of galaxies and halos. In \S 2.2, we 
numerically calculate and project the X-ray emission for a small sample of 
models to verify the predictions of the analytic model, and to examine the 
radial behavior of the twist more closely. In \S 3, we attempt to model the 
large twist observed in NGC 720 (BC94, BC96b), and we present our 
conclusions in \S 4.

\section{CHARACTERISTICS OF X-RAY ISOPHOTE TWISTS}

\subsection{Analytic Projection of Triaxial Ellipsoids}
An ellipsoidal luminosity density $\nu(m^2)$ with constant axis ratios 
$c/a$ and $b/a$ depends only on the ellipsoidal coordinate 
$m^2 \equiv X^2/a^2 + Y^2/b^2 + Z^2/c^2$ (hereafter, we will set $a=1$ for
simplicity). When it is projected in the spherical-polar coordinate 
direction ($\theta , \phi$), with the observer's $z$-axis along the line of 
sight, its surface brightness can be expressed as a one-dimensional 
integral of the radial density profile,
\[
I(x,y)=\frac{2}{\sqrt{f}}\int_0^\infty \nu (u^2 + m_s^2) du , \quad {\rm where}
\]
\begin{eqnarray}
f \equiv \sin^2 \theta (\cos^2\phi + \frac{\sin^2\phi}{b^2})+\frac{\cos^2\theta}{c^2}, \qquad
m_s^2 \equiv \frac{1}{f}(A x^2 + B x y + C y^2), \nonumber \\
A \equiv \frac{\cos^2\theta}{c^2}(\sin^2\phi+\frac{\cos^2\phi}{b^2})+\frac{\sin^2\theta}{b^2}, \qquad
B \equiv \cos\theta \sin 2\phi (1-\frac{1}{b^2})\frac{1}{c^2}, \nonumber
\end{eqnarray}
\begin{equation}
{\rm and} \quad C \equiv (\frac{\sin^2\phi}{b^2} + \cos^2\phi)\frac{1}{c^2}.
\end{equation}
For $\theta=0^\circ$, the line of sight is along the short axis; for 
$\theta=90^\circ, \phi=0^\circ$, it is along the long axis; and for 
$\theta=90^\circ, \phi=90^\circ$, it is along the intermediate axis. The 
surface brightness $I$ is also ellipsoidal, with a constant axis ratio $q$
and a constant minor axis position angle $\psi$ from the projection of the 
short axis. If the galaxy rotates about its short axis, $\psi$ is 
equivalent to the kinematic misalignment angle. The axis ratio and the 
position angle are
\[
q^2 = \frac{A+C-\sqrt{B^2+(A-C)^2}}{A+C+\sqrt{B^2+(A-C)^2}},\quad {\rm and}
\]
\begin{equation}
\tan{2 \psi} = \frac{B}{A-C} = \frac{T \cos \theta \sin 2\phi}{T (\cos^2\theta \sin^2\phi-\cos^2\phi)+\sin^2\theta}.
\end{equation}
Equations 1 and 2 were derived by Binney (1985). The axis ratio 
$q(b,c,\theta,\phi)$ is a function of the projection angles and both axis 
ratios of the triaxial ellipsoid, but the position angle 
$\psi(T,\theta,\phi)$ is a function of only the triaxiality parameter 
$T \equiv(1-b^2)/(1-c^2)$ and the projection angles (Franx, Illingworth, \& 
de Zeeuw 1991). If we now consider two triaxial distributions whose 
three-dimensional axes are intrinsically aligned, but which have different 
triaxialities ($T_1 < T_2$), they are misaligned in projection by the angle
$\Delta \psi$, given by
\[
\sin 2 \Delta\psi = \frac{(T_2-T_1)\sin^2\theta\cos\theta\sin 2\phi}{D_2 D_1}, \quad {\rm where}
\]
\begin{equation}
D_i^2 \equiv T_i^2\cos^2\theta\sin^2 2\phi + [T_i(\cos^2\theta\sin^2\phi-\cos^2\phi)+\sin^2\theta]^2.
\end{equation}
If the two distributions have the same triaxialities ($T_1=T_2$), then the 
misalignment angle is zero ($\Delta \psi = 0$); and if one of the 
distributions is oblate ($T_1=0$), then the misalignment angle is equal to 
the other distribution's kinematic misalignment angle 
($\Delta \psi = \psi_2$).

These expressions for the misalignment angle are applicable to the case of 
a luminous galaxy embedded in a dark halo, where the two distributions have 
differing triaxialities, labeled by $T_*$ and $T_{\rm d}$. The projected 
mass will not be aligned with the projected light, an effect which can be 
important in gravitational lens systems (Keeton, Kochanek, \& Seljak 1997). 
To demonstrate the amplitude of the misalignment for this simple model, 
Figure 1 shows the misalignment angle as a function of the viewing angles, 
$\Delta \psi (\theta, \phi)$, for $T_*=0.25$ and $T_{\rm d}=0.75$. Large 
misalignment angles ($\Delta \psi \gsim 45^\circ$) occur when the 
distribution is viewed near the major axis 
($\sin^2 \phi \lsim 0.05 \cdot \Delta T / T_*$) at intermediate latitudes 
($T_* \lsim \sin^2\theta \lsim T_{\rm d}$).

Figure 2 shows the mean misalignment angle, $\langle \Delta \psi \rangle$, 
and its dispersion, $\sigma_{\Delta \psi}$, averaged over all viewing 
angles ($\theta,\psi$), for all possible triaxialities ($T_*, T_{\rm d}$).
The dispersion $\sigma_{\Delta \psi}$ is comparable to the mean, and
$\langle \Delta \psi \rangle$ increases smoothly with $\Delta T$, so that 
the largest mean misalignments occur when the system consists of a very 
oblate spheroid and a very prolate spheroid. Figure 2 also illustrates the 
misalignment angles expected for a naive model of the shapes of galaxies 
and their halos.  If we assume a luminous galaxy with rough shape limits 
taken from model Ia of Franx, Illingworth, \& de Zeeuw (1991), and embed it 
in a dissipationless dark halo whose shape is taken from the models of 
Dubinski \& Carlberg (1991), we find large typical misalignments, 
$\langle \Delta \psi \rangle \simeq 18^\circ \pm 20^\circ$. If we use the 
model for a dissipational halo from Dubinski (1994), the typical 
misalignment angle drops to 
$\langle \Delta \psi \rangle \simeq 7^\circ \pm 11^\circ$. In both cases, 
misalignments much larger and much smaller than the mean are common. Thus, 
for both galaxy formation models, the projected mass will for some viewing 
angles show a substantial position angle twist from small radii (where the 
luminous matter is dominant) to large radii (where the dark matter is 
dominant).  If the X-ray isophotes show the same asymptotic axes as the 
projected mass, $\langle \Delta \psi \rangle$ will also describe the 
typical asymptotic X-ray misalignment angles. A full numerical model is 
required to confirm this assumption and to determine where in radius the 
isophote twist occurs.
\vspace{0.2in}

{\plotfiddle{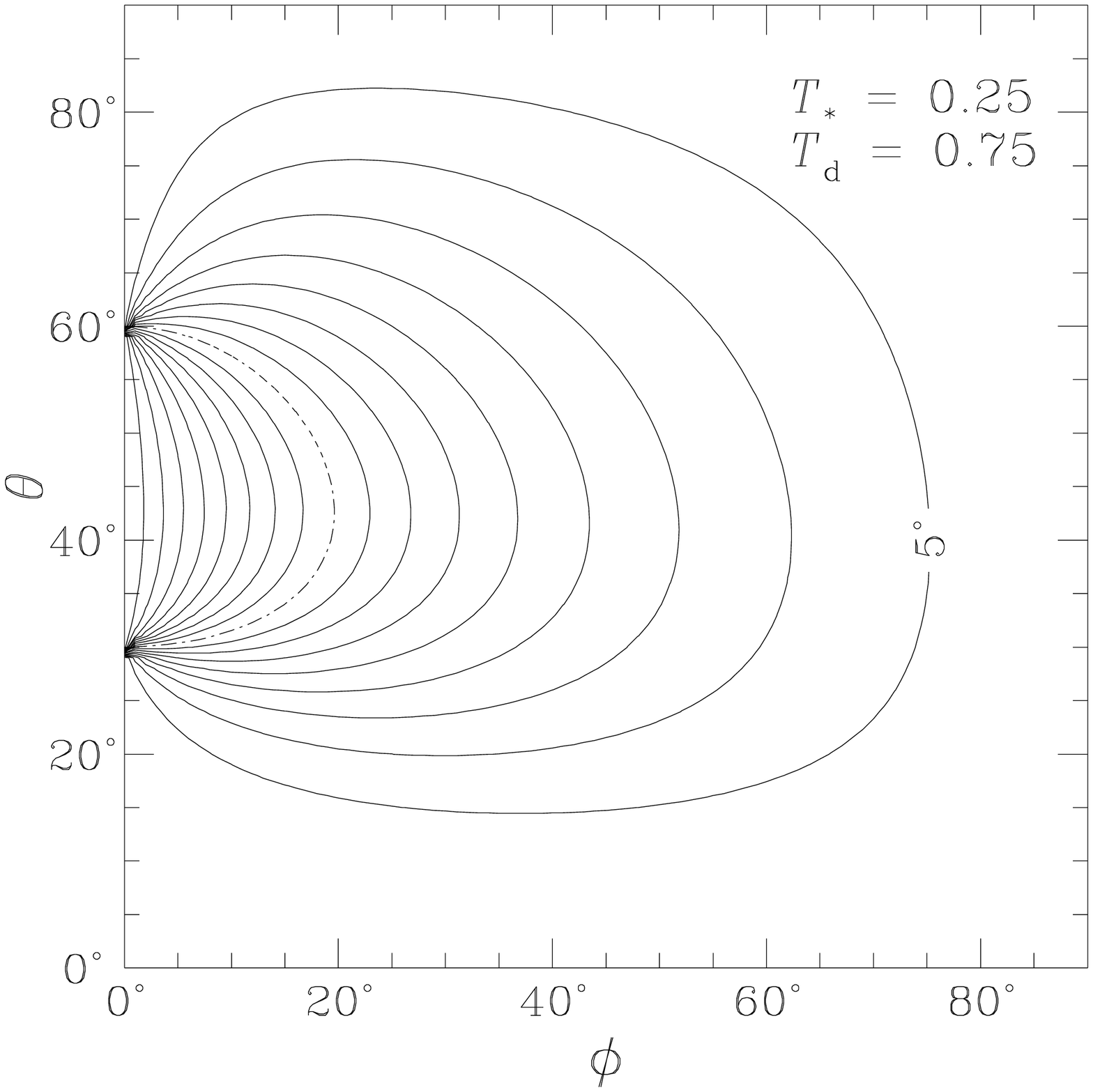}{2.2in}{0}{33.5}{33.5}{125}{-66}}

{\center \small {\bf F{\scriptsize \bf IG}. 1.---} The misalignment angle 
$\Delta \psi$ of two triaxial bodies, for the spherical-polar viewing 
angles $\phi$, $\theta$. The contours are spaced in $5^\circ$ increments
from $5^\circ$ to $85^\circ$, and the solution for 
$\Delta \psi = 45^\circ$ is highlighted by a dot-dash contour.}

\vskip 0.5 cm

\vspace {0.5in}

{\plottwo{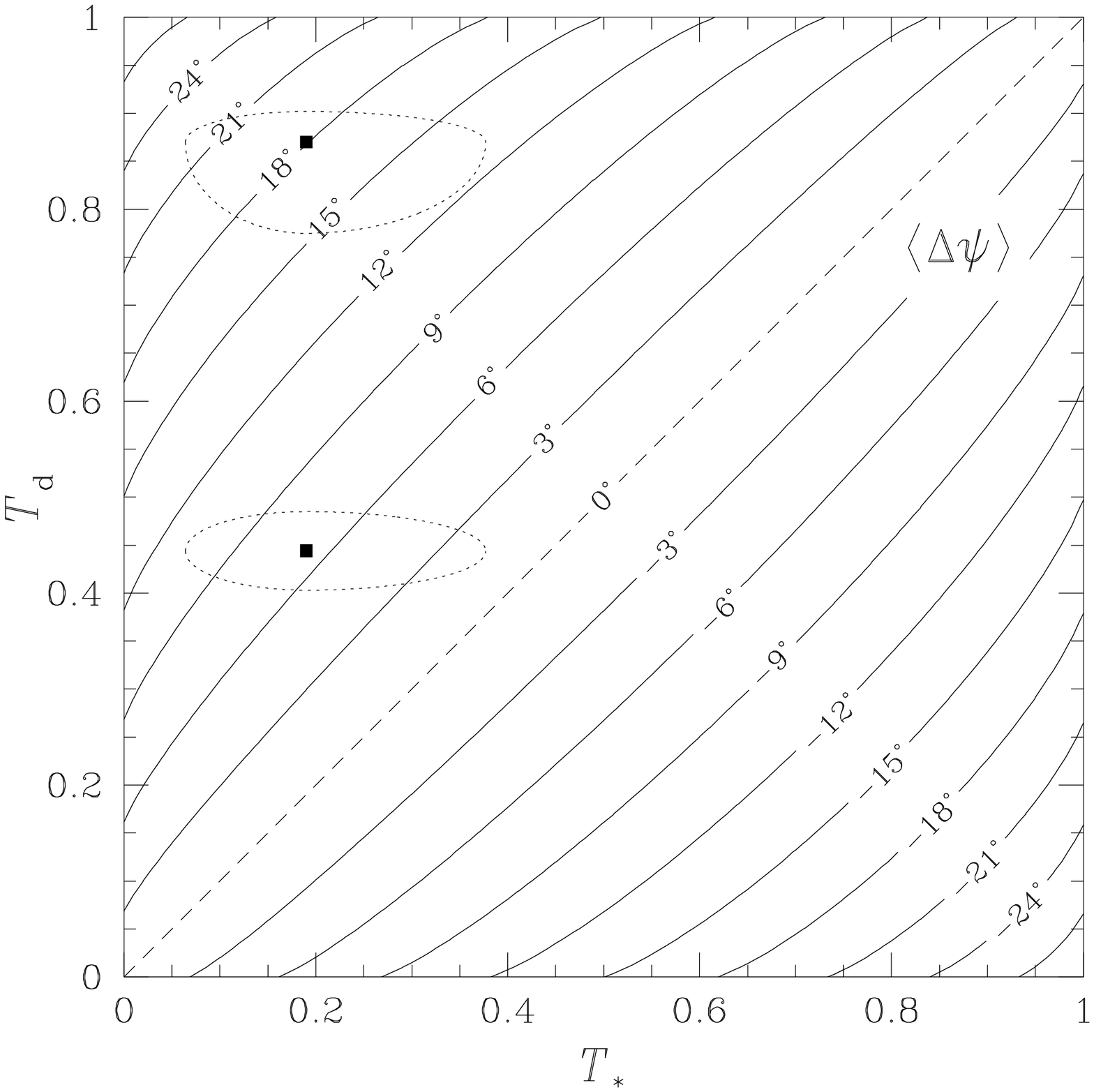}{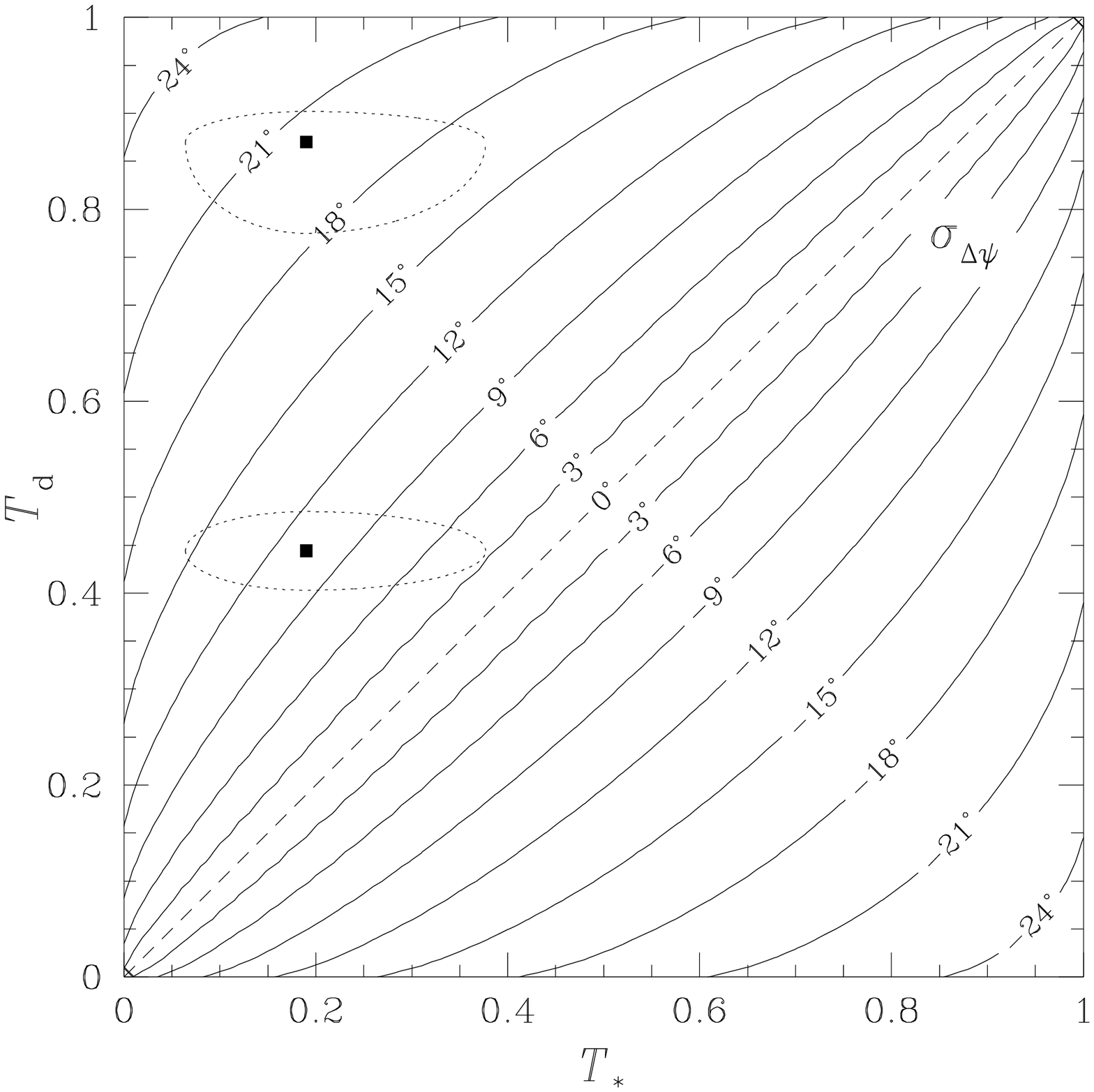}}

{\center \small {\bf F{\scriptsize \bf IG}. 2.---} Contours of the mean 
misalignment angle $\langle \Delta \psi\rangle$ (left) and its dispersion 
$\sigma_{\Delta \psi}$ (right) between two distributions of triaxiality
$T_*$ and $T_{\rm d}$. As an aid to the eye, rough limits from the 
literature for stellar and halo triaxialities ($T_*, T_{\rm d}$) are 
shown for a model of galaxy formation which includes dissipation (bottom) 
and one which does not (top); the squares represent the median values, 
and the dotted curves show the ``one-$\sigma$'' limits.
}

\vskip 0.5 cm

\subsection{Numerical Projection of the X-Ray Emission}
We model the galaxy as a triaxial ``$\eta$-model'' distribution of stars 
(Dehnen 1993; Tremaine et al. 1994) embedded in a triaxial softened 
isothermal dark matter halo potential, with the axes aligned. The total 
density, $\rho ({\bf r}) = \rho_*(m_*^2) + \rho_{\rm d}(m_{\rm d}^2)$,
consists of two ellipsoids of the form
\begin{equation}
\rho_{\rm d}(m_{\rm d}^2) = \frac{\rho_0}{b_{\rm d} c_{\rm d}} \frac{s_{\rm d}^2}{s_{\rm d}^2 + m_{\rm d}^2} \qquad {\rm and} \qquad
\rho_*(m_*^2) = \frac{\eta M_*}{4\pi b_* c_*}\frac{s_* m_*^{\eta-3}}{(s_*+m_*)^{(\eta+1)}}, \quad 0 < \eta \leq 3.
\end{equation}
An ideal, completely-ionized, single-temperature, non-rotating gas in 
hydrostatic equilibrium has a density
$\rho_{\rm g}({\bf r}) = \rho_{\rm g0} \exp [-\mu m_{\rm p}\Phi({\bf r})/k_{\rm B}T_{\rm g}]$
and an X-ray emissivity
\begin{equation}
\nu_{\rm X} ({\bf r}) = \alpha (\lambda_{\rm X}, T_{\rm g}) \rho_{\rm g}^2({\bf r}) = \nu_{\rm X0} e^{-2 \mu m_{\rm p}\Phi({\bf r})/k_{\rm B}T_{\rm g}}
\end{equation}
(Cavaliere \& Fusco-Femiano 1978), where $\mu m_{\rm p}$ is the mean mass per 
particle, the emissivity constant $\alpha$ (when convolved with the detector's
spectral response) is weakly dependent on the gas temperature $T_{\rm g}$, 
and the potential $\Phi({\bf r})$ is evaluated numerically. The asymptotic 
profile is $\nu_{\rm X}(r) \propto r^{-6 \beta}$, where 
$\beta \equiv 4 \pi G \rho_0 s_{\rm d}^2 \mu m_{\rm p}/ 3 k_{\rm B} T_{\rm g}$.
We integrate the X-ray emission along the line of sight to find the surface 
brightness, $I_{\rm X}(x,y) = \int_{-\infty}^\infty \nu_{\rm X}(X,Y,Z) dz$,
and we find the axis ratios $q(a)$ and the minor axis position angles $\psi(a)$ 
of the X-ray isophotes by finding the maximum of $I_{\rm X}$ at each semi-major
axis $a$.

As shown by Buote \& Canizares (1994, 1996a),
for a gas
with any arbitrary temperature profile,
the three-dimensional shape of its X-ray emissivity,
$\nu_{\rm X}$, is the same as that of the three-dimensional potential, $\Phi$, 
in which it lies (i.e., $T_{\rm X} = T_\Phi$). Although the potential is rounder
than the mass, and its axis ratios have a different radial dependence, the 
projected mass and the projected potential are aligned because they must satisfy
the two-dimensional Poisson equation. 
For an isothermal gas, we find that the X-ray emission 
$\nu_{\rm X}$ is also roughly aligned with the projected mass ($\sim 1.5^\circ$ 
typical difference), and thus confirm our assumption that equation 3 and Figure 
2 give an accurate asymptotic representation of X-ray isophote twists.  
Furthermore, by treating the X-ray axis ratios $b_{\rm X}$ and $c_{\rm X}$ as 
those of a perfect triaxial ellipsoid of X-ray emissivity, we can use the 
analytic expressions (1-2) to find the position angle $\psi_{\rm X}(a)$ at any 
radius to an accuracy of $\sim 1^\circ$, and the projected axis ratio 
$q_{\rm X} (a)$ to an accuracy of $\sim 0.01$. 

Figure 3 illustrates in detail the X-ray twists of a realistic galaxy with both 
stars and a dark halo, for several viewing angles. The stellar component is 
specified by a Hernquist (1990) model ($\eta = 2$ in equation 4). 
The relative masses and scale radii of the halo and the stars are set so that
the rotation curve along the long axis is flat. The scale radius ratio is 
$s_*/s_{\rm d} \simeq 1.8$. The mass ratio of the dark matter and stellar 
components within the stellar effective radius ($R_{\rm eff} \simeq 1.8 s_*$) 
is thus $\sim 1$, and within $8 R_{\rm eff}$ is $\sim 6$. The surface density 
profiles of galactic X-ray emission can typically be fit by King-type 
$\beta$-models (see Sarazin \& Bahcall 1977) with $\beta=$ 0.4-0.6 (Forman,
Jones, \& Tucker 1985; Fabbiano 1989), and we set $\beta =$ 0.4.  
The axis ratios of the stellar component are 
chosen to be $b_* = 0.96$ and $c_* = 0.75$, corresponding to $T_* = 0.18$. The 
axis ratios of the halo in the dissipationless scenario are $b_{\rm d}=0.58$ 
and $c_{\rm d}=0.42$, corresponding to $T_{\rm d}=0.81$; and in the 
dissipational scenario they are $b_{\rm d}=0.85$ and $c_{\rm d}=0.62$, 
corresponding to $T_{\rm d}=0.45$. The minor axis position angle of the X-ray 
isophotes, $\psi_{\rm X}$, matches roughly the position angle of the light, 
$\psi_*$, at small radii, and that of the halo, $\psi_{\rm d}$, at large radii, 
with a gradual transition at intermediate radii. The radial behavior (e.g., the 
location and sharpness) of this transition is insensitive to the gas temperature 
parameter $\beta$, and is somewhat sensitive to the viewing angles, the stellar 
exponent $\eta$, the scale radii, and the mass ratio.  The half-way point of the 
twist generally occurs at semi-major axis $a \sim$ 0.1-0.4$R_{\rm eff}$ for the 
dissipationless halo model, and at $\sim$ 0.4-0.6$R_{\rm eff}$ for the 
dissipational halo model. As predicted by the analytic description, there is a 
large scatter in the twist amplitude, depending on the viewing angles. For the 
dissipationless halo, the median asymptotic twist is 
$\Delta \psi_{\rm X} \sim 15^\circ$, and our small sample shows twists as small 
as $3^\circ$ and as large as $45^\circ$. For the dissipational halo, the median 
twist is $\sim 6^\circ$, and twists from $1^\circ$ to $20^\circ$ are seen. Note 
that although the halo begins to dominate the total mass only outside 
$R_{\rm eff}$, the X-ray isophote twist can occur much closer to the core. This 
is because the triaxiality of the X-ray emissivity $T_{\rm X}$ is not a linear 
function of the axis ratios ($b_{\rm X}$, $c_{\rm X}$), and in this case where 
the stellar potential is nearly oblate, the small amount of additional 
flattening that is produced by the halo along the intermediate axis is 
sufficient to alter the triaxiality (and thus the isophote position angle) 
considerably. With our numerical model, we reappraise the limits from Figure 2 
and find that at $3 R_{\rm eff}$, the mean misalignment angle for the 
dissipationless model is 
$\langle \Delta \psi_{\rm X} \rangle \simeq 16^\circ \pm 19^\circ$, and for the 
dissipational model is $\simeq 5^\circ \pm 8^\circ$.

The axis ratios of the X-ray isophotes are less strongly affected by triaxiality
differences and parameter adjustments. For the dissipationless halo, 
$q_{\rm X} \sim$ 0.85-0.9 at $R_{\rm eff}$ for most viewing angles, and for the 
dissipational halo, $q_{\rm X} \sim$ 0.9-0.95. Figure 4 shows X-ray isophote 
contours in the central regions of the galaxy for both halo models and some 
different viewing angles. Note that the larger twists occur for rounder 
isophotes, whose position angles are more difficult to determine. Although it is
not apparent from the figure, the position angle of the X-ray isophotes 
asymptotically approaches that of the projected halo mass.  In many cases we 
will not be able to resolve the X-ray twist, but we can still recognize its 
presence and measure its amplitude by observing the misalignment between the 
stellar and the outer X-ray isophotes.

As a further complication, motivated by observations of the galaxy NGC 720 
(BC97), we add a contribution from discrete X-ray sources 
whose luminosity distribution is directly proportional to the stellar mass 
distribution $\rho_*$, and whose total emission is 20\% of the gaseous X-ray 
emission within $5 R_{\rm eff}$. This discrete component significantly modifies 
the triaxiality of the X-ray emission, and causes the X-ray twist to remain 
relatively flat to a much larger radius (see Figure 3). The character of the 
twist also becomes much more sensitive to the galaxy model parameters, including
the gas temperature $\beta$, once we include a discrete component.

{\plotone{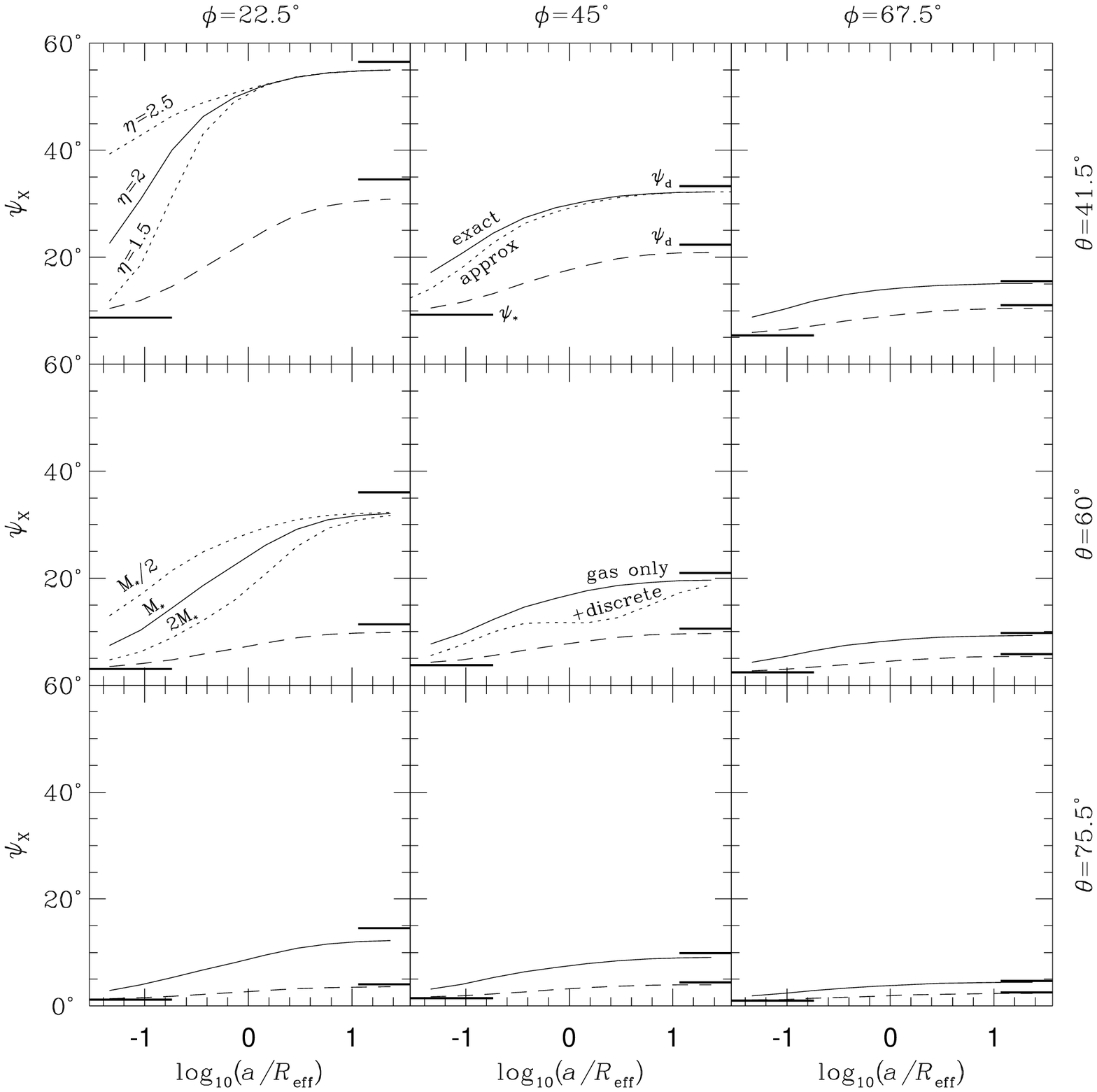}}
{\center \small {\bf F{\scriptsize \bf IG}. 3.---} The minor-axis position angle
of the X-ray isophotes of a model galaxy (measured relative to the projected 
kinematic axis), as a function of the semi-major axis, for a gridded sample of 
viewing angles. The solid curves correspond to the dissipationless model, and 
the dashed curves to the dissipational model. Some special cases are also 
displayed for the dissipationless model (dotted curves). The center plot shows a
case with a discrete source component added. The center left plot shows cases 
with different stellar component masses, $M_*$. The top left plot shows cases 
with different values for the stellar exponent, $\eta$. The top center plot 
shows the analytic approximation for the twist angle. The stellar and halo 
misalignment angles ($\psi_*$ and $\psi_{\rm d}$) are marked (see top center 
plot for labeling).
}

{\plotone{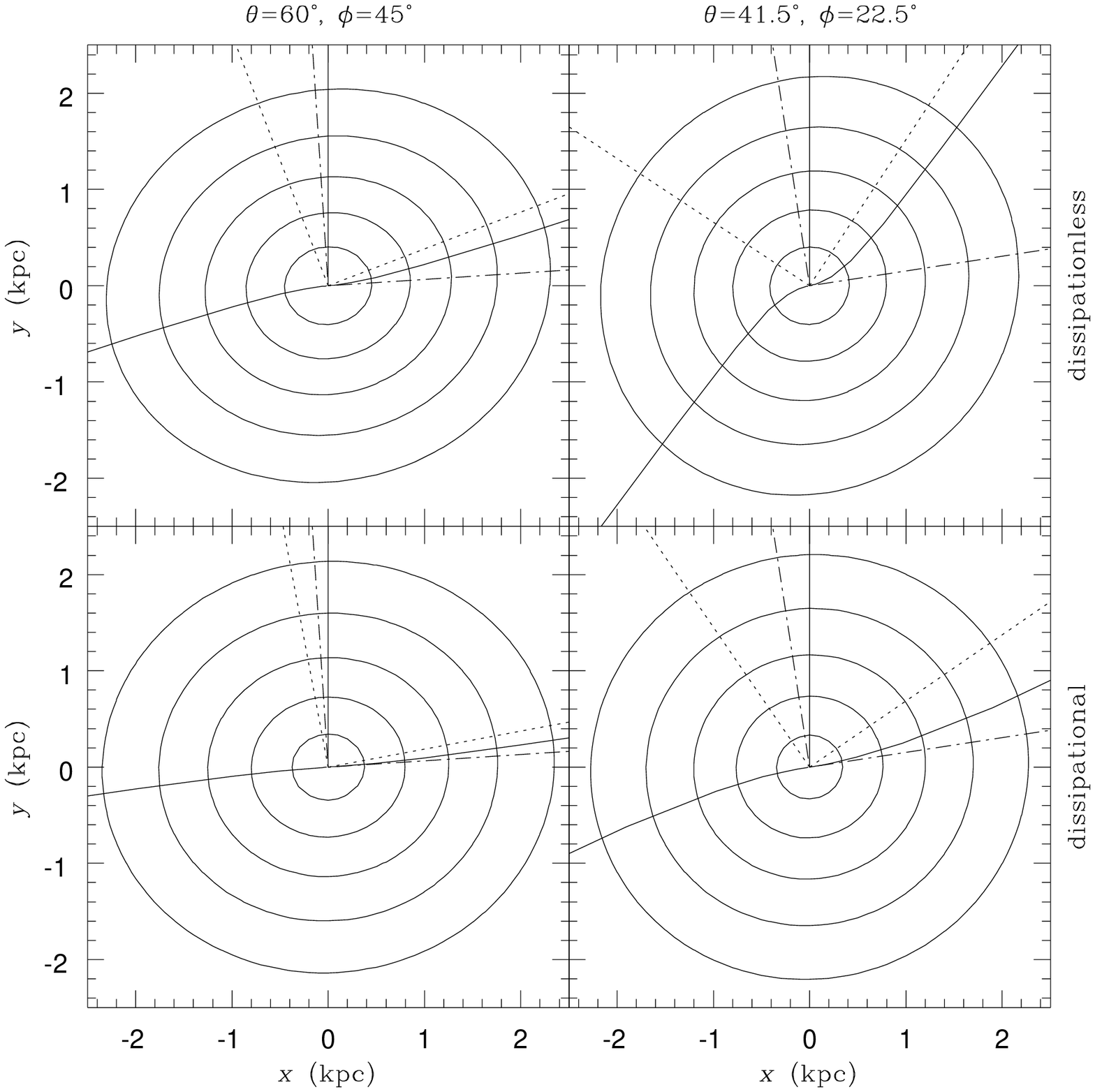}}
{\center \small {\bf F{\scriptsize \bf IG}. 4.---} X-ray isophotes of two model 
galaxies, for different viewing angles. The contours are logarithmically spaced, 
with the central levels omitted for clarity. The major and minor axes of the 
projected mass are marked for the halo (dotted axes) and for the light 
(dot-dash axes). The kinematic axis $\Psi_{\rm k}$ is marked by a vertical solid 
line. The (twisting) X-ray major axis is shown by a solid line. The top two 
cases correspond to a model with a halo from the dissipationless scenario, while 
the bottom cases are from the dissipational scenario. The stellar effective 
radius is $\simeq$ 5.4 kpc.
}

\vskip 0.5 cm

\section{A KNOWN EXAMPLE: NGC 720}

Finally, we consider the particular case of NGC 720, an isolated E4 galaxy with 
bright X-ray emission (Forman, Jones, \& Tucker 1985), and no evidence that
environmental effects distort its X-ray isophote shapes significantly (BC96b).
According to Binney, Davies, \& Illingworth (1990), the stellar misalignment 
angle $\psi_*$ is consistent with zero. Buote \& Canizares (BC94, BC96b, BC97)
have carefully studied and modeled NGC 720 based on data from {\it ROSAT}'s PSPC 
and HRI, and {\it ASCA}. They found that the X-ray isophotes are significantly 
elongated ($q_{\rm X} \sim$ 0.75 for 
$a \simeq$ 60$^{\prime \prime}$-110$^{\prime \prime}$, where 
$1^{\prime \prime} \sim 0.1 h_{80}$ kpc), with some indication that the 
ellipticity decreases at large radii ($q_{\rm X} \sim$ 0.8-0.9 for 
$a \gsim$ 120$^{\prime \prime}$). There is a significant X-ray isophote twist 
between the the inner and the outer isophotes, where the X-ray major axis 
position angle $\Psi_{\rm X}$ is consistent with being aligned with the optical 
angle $\Psi_*$ at small radii ($\Psi_{\rm X} \simeq 142^\circ$ for 
$a \lsim$ 60$^{\prime \prime}$), and twists to a different angle at large radii 
($\Psi_{\rm X} \simeq 114^\circ$ for $a \gsim 90^{\prime \prime}$). The emission 
comes primarily from hot isothermal gas, with a lesser component ($\sim 20$\%) 
from X-ray binaries (BC97). 
BC94 demonstrated that the shape alone of the X-ray emission 
necessitates the presence of an extended dark matter halo, since the potential 
produced by the stellar distribution is too round to produce the high 
ellipticity observed for the X-ray emission. 
BC96b suggested triaxiality of the 
halo as a possible culprit for the X-ray isophote twist, but they were unable to 
reproduce the twist using a simple triaxial model in which the intrinsic axis 
ratios vary with radius according to a power law.

We reinvestigate the hypothesis that triaxiality is responsible for the observed 
X-ray isophote twist by searching for a reasonable galaxy model whose stellar 
and halo components produce a radially-varying axis ratio $q_{\rm X} (a)$ and 
minor axis position angle $\psi_{\rm X} (a)$ consistent with those reported for 
NGC 720. Taking advantage of the fact that we can use the analytic equations 
(1-2) to find good approximations for $q_{\rm X}$ and $\psi_{\rm X}$, we try to 
find reasonable values for the parameters 
$\vec{\xi} = (T_*, c_*, T_{\rm d}, c_{\rm d}, \theta, \phi)$ which will give us 
the observed data $\vec{\delta} = (\psi_*, q_*, \psi_{\rm X}, q_{\rm X})$. We 
use a Bayesian statistical technique to find the most probable solution, and 
examine first the case with a dissipationless halo.  For our priors, we use 
Gaussian distributions of the parameters ($c_*, c_{\rm X}, T_*, T_{\rm X}$), 
where $\langle c_* \rangle = 0.75 \pm 0.2$, 
$\langle c_{\rm X} \rangle = 0.79 \pm 0.08$, 
$\langle T_* \rangle = 0.2 \pm 0.2$, and
$\langle T_{\rm X} \rangle = 0.7 \pm 0.3$. For the data, we set
$\psi_* = 0^\circ \pm 10^\circ$, $q_*=0.60 \pm 0.05$,
$\psi_{\rm X} = 27^\circ \pm 8^\circ$, and $q_{\rm X}=0.75 \pm 0.07$,
and we assume that the kinematic axis is intrinsically aligned with the short 
axis.

Figure 5 illustrates the solutions, showing the total posterior probability of
the parameters given the data, 
$P(\xi_i| \vec{\delta}) = \Sigma_{j \neq i} P(\vec{\xi} | \vec{\delta})/\Sigma_j P(\vec{\xi} | \vec{\delta})$,
where $P(\vec{\xi} | \vec{\delta}) = P(\vec{\xi}) P(\vec{\delta} | \vec{\xi})$.
The maximum likelihood occurs at ($\theta \simeq 67^\circ$, 
$\phi \simeq 42^\circ$, $c_* \simeq 0.51$, $c_{\rm X} \simeq 0.71$, 
$T_* \simeq 0.17$, $T_{\rm X} \simeq 0.96$), corresponding to 
$\psi_*\simeq 2.3^\circ$, $\psi_{\rm X}\simeq 21.5^\circ$, $q_* \simeq 0.62$,
$q_{\rm X}\simeq 0.81$. This solution is a marginal fit to the data given the 
errors in the axis ratios and position angles of the X-ray isophotes. The 
solution for the stellar shape is reasonable --- although the optical 
ellipticity of NGC 720 varies considerably with radius, there is no significant 
isophote twist (Jedrzejewski, Davies, \& Illingworth 1987; Capaccioli, Piotto,
\& Rampazzo 1988; Peletier et al. 1990; Nieto et al. 1991; Sparks et al. 1991),
indicating either that the stellar distribution is very oblate, or that it is 
triaxial and we are observing it nearly ``edge-on''. The X-ray emissivity shape 
corresponds to a flat, prolate halo with $c_{\rm d} \simeq 0.34$, 
$T_{\rm d} \simeq 1.0$, which is consistent with the models of BC94, where 
$c_{\rm d} \sim$ 0.2-0.5, assuming an edge-on viewing angle and either an oblate 
or a prolate halo. If we use priors on the halo shape from the dissipational 
collapse scenario, it is significantly more difficult to fit the data.

{\plotfiddle{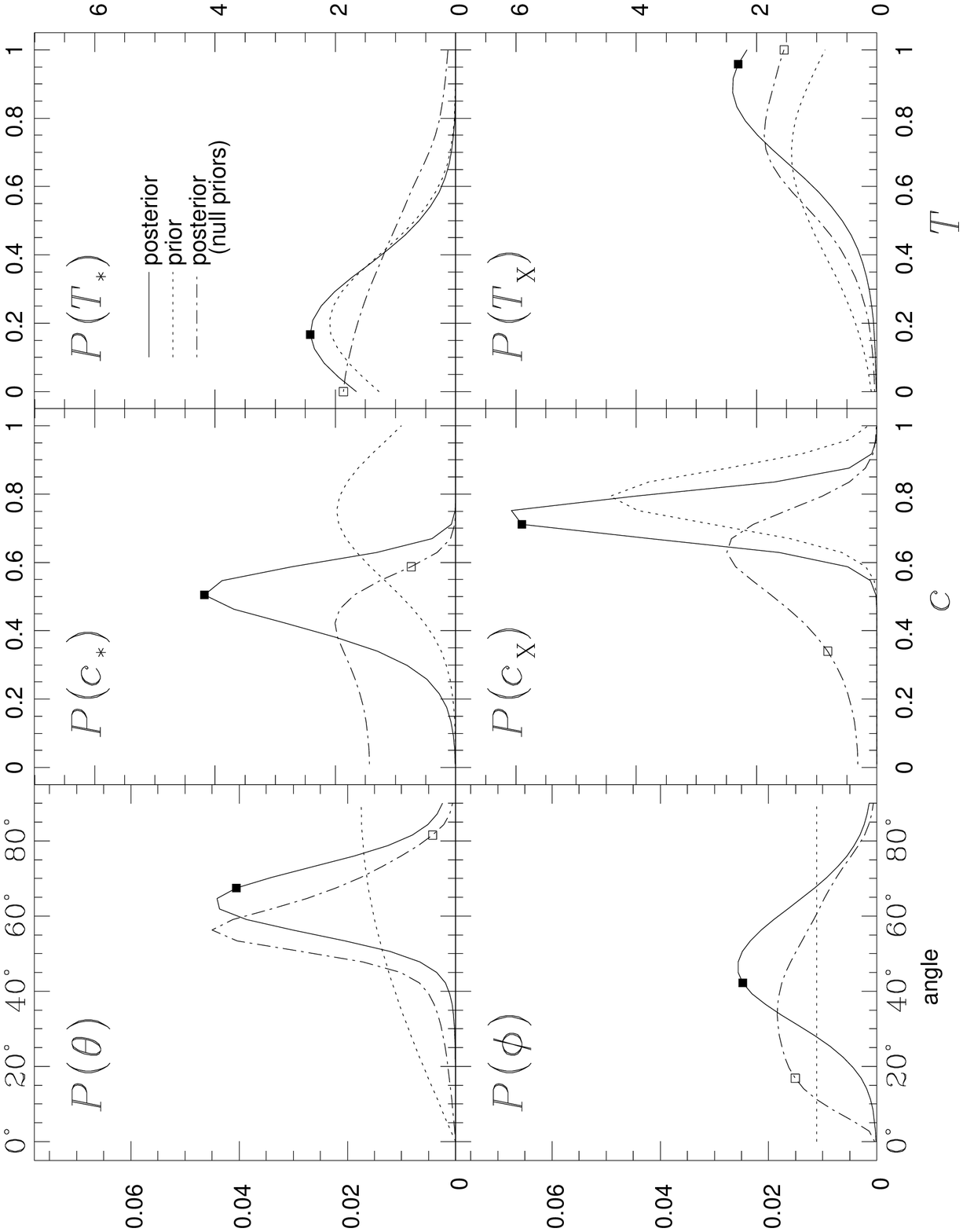}{4.8in}{-90}{60}{60}{-17}{351}}

{\center \small {\bf F{\scriptsize \bf IG}. 5.---} The projected likelihood 
distribution for the model parameters to fit the data for NGC 720. The prior 
distributions are plotted, as are the posterior distributions which included the 
priors, and the posterior distributions which did not include the priors 
(excepting the prior on $\theta$). The squares mark the maximum likelihood 
solution.
}

\vskip 0.5cm

We next use the maximum likelihood model to see if it can reproduce the shapes 
and orientations of the X-ray isophotes in detail. Since the stellar effective 
radius is $R_{\rm eff} \simeq 52^{\prime \prime}$ (Burstein et al. 1987), we 
take the scale radius to be $s_* = 28\parcs6$. After BC94, we set the 
gas temperature to give a radial profile exponent of $\beta=0.5$.  Since BC94 
find that inside 450$^{\prime \prime}$, the ratio of the masses of the halo and 
the stars $\gsim 4$ at 90\% confidence, with no upper bound, we use a model with 
a mass ratio of $\sim 0.7$ inside $R_e$ and $\sim 7$ inside 
$450^{\prime \prime}$ ($\sim 9 R_{\rm eff}$). From the fits of BC94, we take a 
typical halo scale radius to be $s_{\rm d}=59\parcs6$. Figure 6 shows 
the model X-ray isophotes, and Figure 7 compares the radial variation of the 
position angle and axis ratio of its projected X-ray emission to the 
data\footnote{Note that our models measure $\Psi_{\rm X}$ and $q_{\rm X}$ along isophotes,
while the data measure them using a moment method (compare $\epsilon_{\rm M}$ and $\epsilon_{\rm iso}$
from BC94);
these do not differ significantly unless there are large radial gradients.}.
Although the twist and the ellipticity at large radii are marginally compatible 
with the data, the twist transition is too gradual and too close to the core.
Despite much experimentation with the model parameters, we could not 
significantly improve our fit to the transition. By adding a discrete 
source component to the X-ray emission, we can push the transition radius 
further outward, but it becomes far more difficult to fit the amplitude 
of the twist.
The effects are similar if we 
compute the twist with the moment method (BC94),
or if we allow for substantial rotation of the X-ray emitting gas
in the central regions. We thus qualitatively confirm the tentative 
result of BC96b that it is very difficult for simple models with a radial 
triaxiality gradient to simultaneously fit both the high ellipticity and the large, 
abrupt twist of the observed X-ray isophotes.

{\plotfiddle{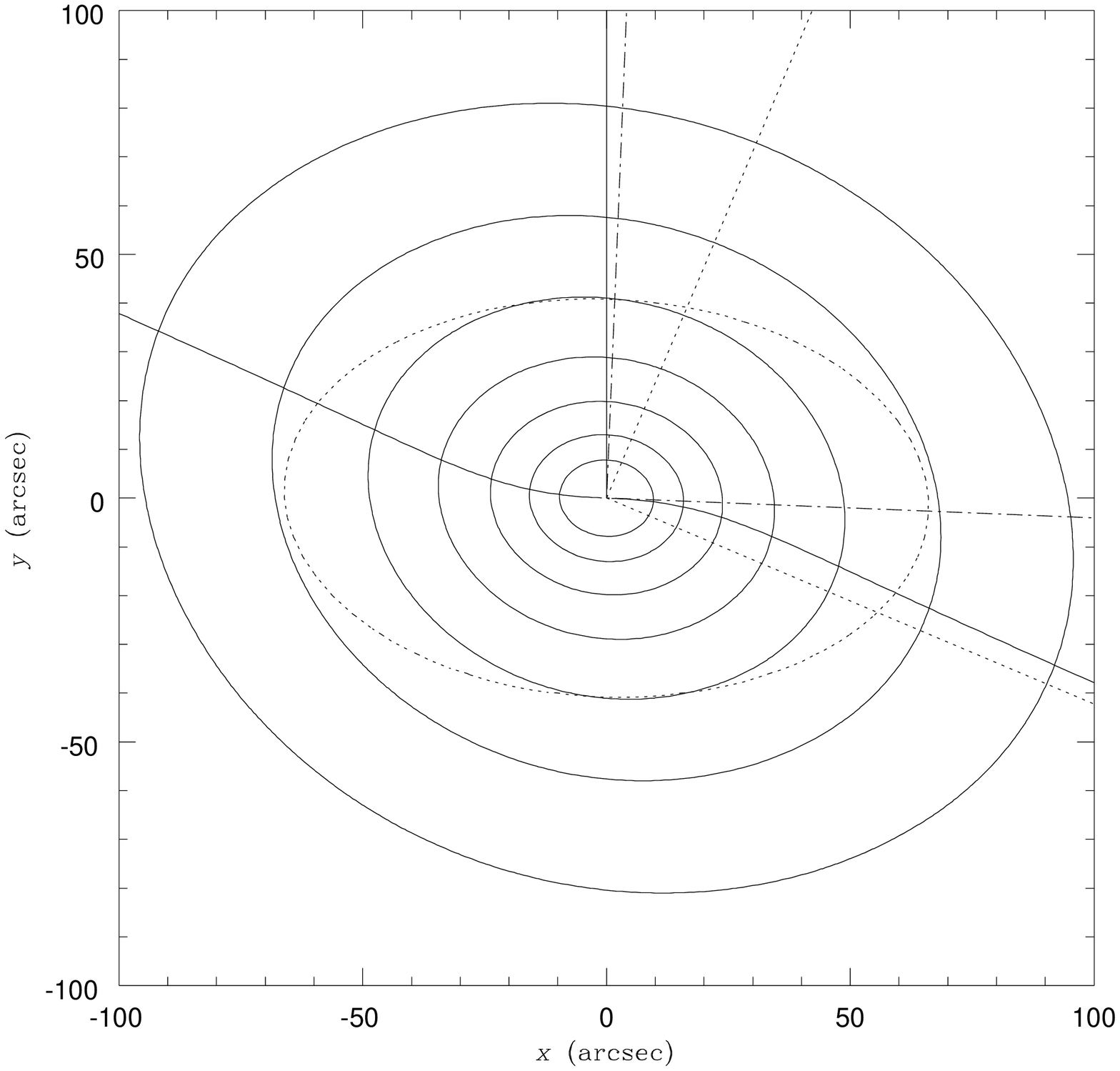}{3.5in}{0}{50}{50}{76}{-90}}

{\center \small {\bf F{\scriptsize \bf IG}. 6.---} X-ray isophotes of our model 
of NGC 720. (Compare BC94 Fig. 5.) The contours are spaced logarithmically by a 
factor of two, with the central levels omitted for clarity. The dotted contour 
shows the stellar surface density isophote at the stellar effective radius. The 
major and minor axes of the projected mass are marked for the halo (dotted axes)
and for the light (dot-dash axes). The kinematic axis $\Psi_{\rm k}$ is marked 
by a vertical solid line. The (twisting) X-ray major axis is shown by a solid 
line.
}

\vskip 0.5cm

{\plottwo{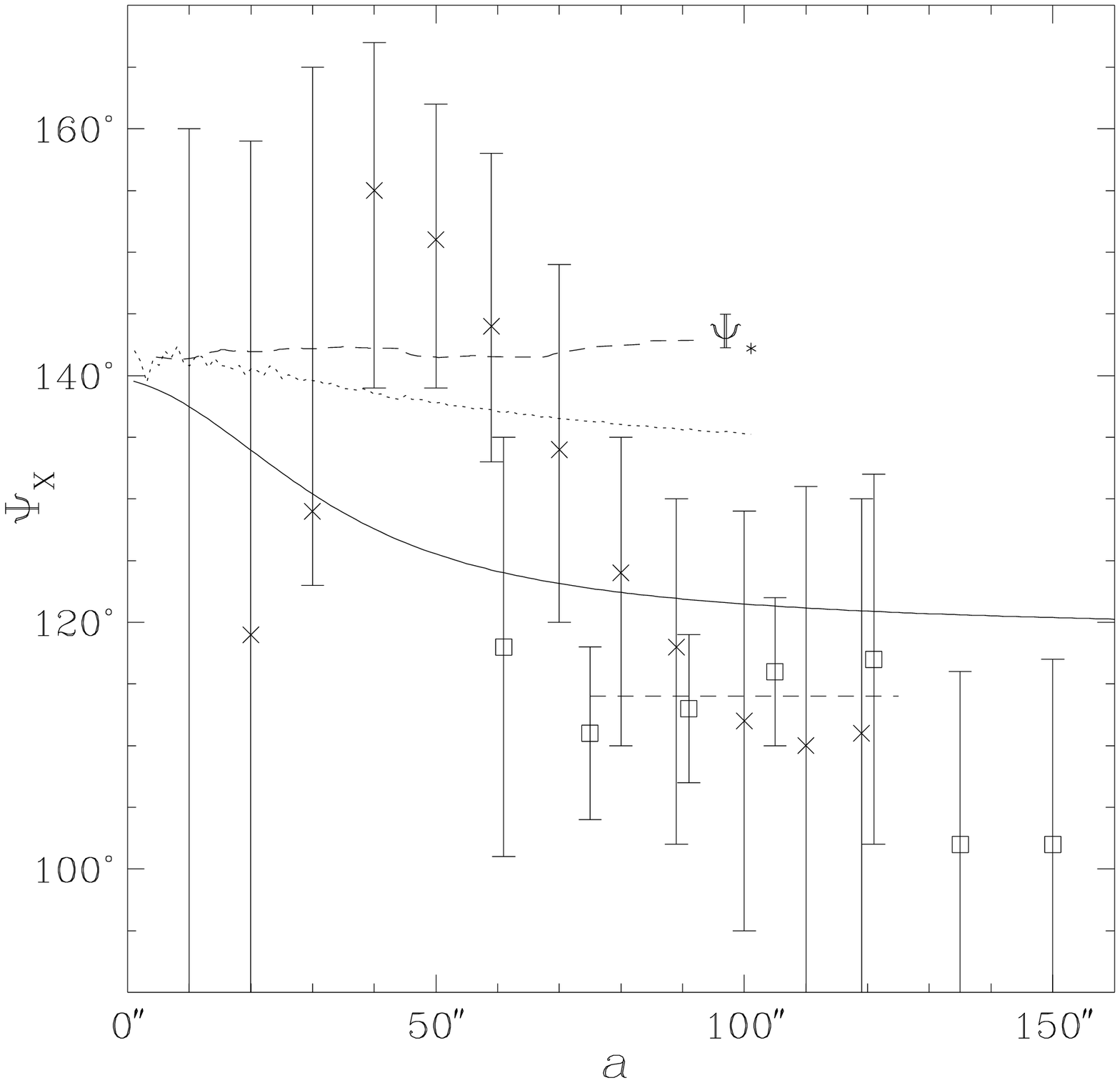}{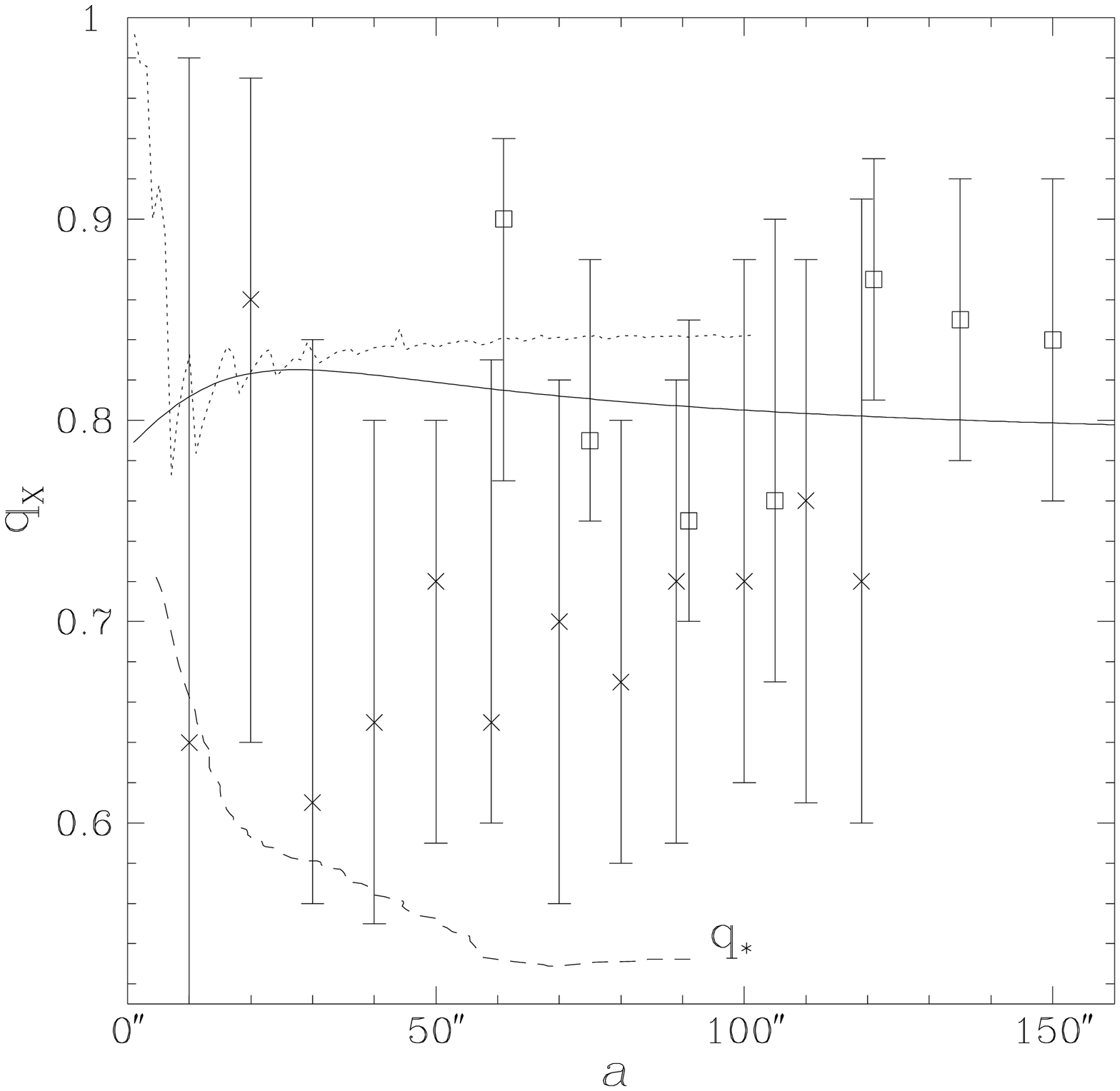}}

{\center \small {\bf F{\scriptsize \bf IG}. 7.---} 
The major axis position angle (left) and the axis ratio (right) of the X-ray 
isophotes as a function of the semi-major axis. Our best-fit model is shown,
where the parameters are calculated on isophotes (solid curves) and using a 
moment method (dotted curves),
and the data are shown with their 90\% confidence limits (for 
clarity, data points at the same radius are offset slightly in radius). The 
{\it PSPC} data are indicated by squares, and the {\it HRI} data by crosses.
The position angle ($\Psi_*$) and the axis ratio ($q_*$) of the stellar 
isophotes are marked by dashed lines 
(taken from Jedrzejewski, Davies \& Illingworth 1987; 
Capaccioli, Piotto, \& Rampazzo 1988; Peletier et al. 1990; Sparks et al. 
1991), as is the approximate position angle of the outer X-ray isophotes.
}

\section{CONCLUSIONS}
Current observational evidence indicates that the stellar parts of elliptical 
galaxies are modestly flattened and close to oblate, while current theories for 
the formation of galaxies suggest that the halos are more flattened and more 
prolate. The extent of this shape difference depends on the interactions between 
baryons and dissipationless dark matter during galaxy formation, but all current 
models imply that the mass distribution of an early type galaxy increases in 
flatness and prolateness with radius.  There is already some evidence from the 
flatness of X-ray isophotes (Buote \& Canizares 1994, 1996a) and from 
gravitational lens models (King \& Browne 1996; Kochanek 1996) that the typical 
mass distribution may be flatter than the luminosity distribution. A difference 
in triaxiality between luminous galaxies and their dark halos is detectable
through misalignments of gravitational lenses with their stellar distributions,
and is a natural explanation for the second shear axis that is necessary to fit 
lensing models, although intrinsic misalignments and external tidal shear 
sources are also viable explanations (Keeton, Kochanek, \& Seljak 1997).  
Such a triaxiality difference also produces X-ray isophote twists (Binney 1978; 
Buote \& Canizares 1996b).

We examined the behavior of X-ray isophote twists using several simple models of 
a luminous galaxy of constant triaxiality embedded in an intrinsically-aligned 
dark halo of a different triaxiality. A simple analytic approximation gives 
accurate estimates of the asymptotic position angles of the X-ray isophotes, but
predicting the detailed radial behavior of the twist requires numerical 
simulations.  For a reasonable model of a galaxy and halo, we find that the 
X-ray isophote position angle makes a gradual transition from the center of the 
galaxy, where it is aligned with the position angle of the optical isophotes, to 
the periphery, where it is aligned with the position angle of the projected halo 
mass.  The ``half-way'' point of the twist occurs well inside the stellar 
effective radius ($\sim$~0.1-0.6 $R_{\rm eff}$), and is not detectable given 
present X-ray resolution limits, but the misalignment of the X-ray isophotes at 
large radii with the stellar isophotes is more easily observable. By examining 
the amplitude of the misalignment for a large population of galaxies, we can in 
principle distinguish between alternative models for the halo shapes.  The very 
prolate halos predicted by dissipationless collapse simulations should produce 
mean misalignments of 
$\langle \Delta \psi_{\rm X} \rangle \simeq 16^\circ \pm 19^\circ$ at 
$\sim 3 R_{\rm eff}$, while the more oblate halos predicted by simulations which 
include baryonic dissipation should produce smaller misalignments 
($\simeq 5^\circ \pm 8^\circ$).  In practice, measurements of these twists will 
be difficult because the twist angles are small, the X-ray isophotes are fairly 
round ($\langle q_{\rm X} \rangle \sim 0.9$), and the number of isolated, nearby 
X-ray galaxies is limited. Our results are not very sensitive to the parameters 
of the model, but the presence of a discrete source component can modify the 
detailed properties of the twist and should be included in any statistical test.
Note that while we scaled our models to match the halo shapes predicted by 
simulations, {\it any} mass distribution with a strong radial triaxiality 
gradient will produce an X-ray isophote twist when observed from the proper 
angles.  For example, the disruption of box orbits by a central stellar cusp can 
cause the central parts of an elliptical galaxy to become more oblate than the 
outer parts due to the longer time scales for scattering and phase mixing of the 
outer orbits (e.g. Merritt 1997b), leaving a twisting signature in the X-ray 
isophote position angles.

We attempted to model the observed X-ray twist of the bright, isolated E4 galaxy 
NGC 720 (BC94, BC96b), whose X-ray isophotes are strongly flattened, and show a 
large, abrupt position angle twist of $\sim 30^\circ$ at 
$\sim$ 1-2 $R_{\rm eff}$.  While a model halo from the dissipationless scenario 
fits the data better than does a halo from the dissipational scenario, it is in
both scenarios difficult to reproduce both the twist behavior and the isophote 
flatness using reasonable parameters for the galaxy and halo.
The simplest explanation for NGC 720 is that the halo and the stars are
intrinsically misaligned. Such an intrinsic 
misalignment may arise naturally for a halo which forms by dissipationless 
collapse, or it may be caused by late-history major mergers.
Whether caused by triaxiality or by intrinsic misalignment, the shapes and
orientations of stellar and X-ray isophotes are important fossil clues to the
formation history of galaxies.

\vspace{1.5cm}
We thank David Buote and Christine Jones for their helpful comments.
CSK is supported by NSF grant AST-9401722 and NASA ATP grant NAG5-4062.

\end{document}